\documentclass[twocolumn,showpacs,preprintnumbers,amsmath,amssymb]{revtex4}

\usepackage{graphicx}
\usepackage{dcolumn}
\usepackage{bm}

\usepackage{epsfig}

\begin{document}

\title{Dynamics of Quintessence with Thermal Interactions}

\author{Dao-jun Liu}
\author{Xin-zhou Li}%
 \email{kychz@shnu.edu.cn}
\affiliation{%
Shanghai United Center for Astrophysics(SUCA), Shanghai Normal
University, 100 Guilin Road, Shanghai 200234,China\\
 Division of
Astrophysics, E-institute of Shanghai Universities, Shanghai
Normal University, 100 Guilin Road, Shanghai 200234,China
}%

\date{\today}

\begin{abstract}
The cosmological dynamics of minimally coupled scalar field that
couple to the background matter with thermal interactions is
investigated in exponential potential. The conditions for the
existence and stability of various critical points as well as
their cosmological implications are obtained. Although we show
that the effects of thermal interaction such as depressing the
equation-of-state parameter of quintessence, is only important at
the early time, the evolution of equation-of-state parameter of
quintessence is manifested. The upper bound is required on the
coupling between quintessence and relativistic relic particles
such as photons and neutrinos.
\end{abstract}

\pacs{98.80.Cq}
\maketitle

Astronomical measurements from Supernovae Ia \cite{riess},
galaxies clustering, for example, Sloan Digital Sky Survey (SDSS)
\cite{Tegmark} and anisotropies of Cosmic Microwave Background
Radiation (CMBR) \cite{Spergel} independently suggest that a large
fraction of the energy density of our Universe is so-called dark
energy, which has negative pressure, or equation of state with
$w=p/\rho<0$, and accelerates the expansion of the Universe. The
origin of the dark energy remains elusive from the point of view
of general relativity and standard particle physics. One candidate
source of this missing energy component is a slowly evolving and
spatially homogeneous scalar field, referred to as "quintessence"
with $w>-1$ \cite{Peebles, Steinhardt, Li} and "phantom" with
$w<-1$ \cite{Caldwell, Carroll,hao1,Sami} respectively. Since
current observational constraint on the equation of state of the
dark energy lies in the range $-1.38<w<-0.82$ \cite{Melchiorri},
it is still too early to rule out any of the above candidates. To
study the global property of the cosmological system containing
dark energy, phase space analysis is proved to be a powerful tool.

Because dark energy redshifts more slowly than ordinary matter or
radiation, there exists an important problem that various
proposals of the dark energy should explain, that is, why the
energy density of matter and dark energy should be comparable at
the present epoch. A form of quintessence called "tracker fields",
whose evolution is largely insensitive to initial conditions and
at late times begin to dominate the universe with a negative
equation of state, was introduced to avoids the above problem
\cite{Zlatev}. Another approach to solve the puzzle is introducing
an interaction term in the equations of motion, which describes
the energy flow between the dark energy and the rest matter
(mainly the dark matter) in the universe. It is found that, with
the help of a suitable coupling, it is possible to reproduce any
scaling solutions.

The effects of thermal coupling between the quintessence field and
the ordinary matter particles has been investigated \cite{Hsu}.
Hsu and Murray set the quintessence field to be a static external
source for a Euclidean path integral depicting the thermal degree
of freedom and let the time-like boundary conditions of the path
integral have a period which is decided by the inverse of
temperature. They show that if in the early universe matter
particles are in thermal equilibrium, quantum gravity
\cite{Kamionkowski} will induce an effective thermal mass term for
quintessence field $\phi$, which takes the form
\begin{equation}\label{eq1}
\left(\frac{\beta}{M_P}\right)^2\phi^2 T^4,
\end{equation}
where $T=\sqrt[4]{\rho_{r0}}/a$ is the temperature of the
universe, $M_P$ the Planck mass scale, $\beta$ a dimensionless
constant, $a$ the scale factor (with current value $a_0=1$) and
$\rho_{r0}$  the energy density of radiation at the present epoch,
and find that even Planck-suppressed interactions between matter
and the quintessence field can alter its evolution qualitatively.
For convenience, Hsu and Murray \cite{Hsu} use an approximation
that the back reaction on matter of quintessence is neglectable.
However, this back-reaction effect, in fact, is significant both
at the early time and late time. In this letter, we investigate
the dynamics of the cosmology with quintessence using the phase
space analysis, which has the above thermal coupling to the matter
in a complete manner. We shall seek the conditions for the
existence and stability of various critical points as well as
their cosmological implications. As will be shown in this paper,
although the effects of thermal interaction, such as depressing
the equation-of-state parameter of quintessence, is only important
at the early time, the evolution of equation-of-state parameter of
quintessence is manifested. The upper bound is required on the
coupling between quintessence and relativistic relic particles
such as photons and neutrinos.

 We first study the phase space of quintessence
with a thermal coupling to ordinary matter particles in spatially
flat FRW cosmological background
\begin{equation}\label{metric}
ds^2=dt^2-a^2(t)d\textbf{x}^2.
\end{equation}
For the spatially homogeneous scalar field minimally coupled to
gravity with thermal interaction (\ref{eq1}), the evolution is
governed by the Klein-Gordon equation
\begin{equation}\label{ddphi}
\ddot{\phi}+3H\dot{\phi}+V'(\phi)=2\left(\frac{\beta}{M_P}\right)^2\phi
T^4,
\end{equation}
where the overdots denote the derivative with respect to cosmic
time and the prime denotes the derivative with respect to $\phi$.
Here the Hubble parameter $H\equiv \dot{a}/a$ is determined by the
Friedmann equation
\begin{equation}\label{FRW}
H^2=\frac{\kappa^2}{3}\left[\rho_m+\frac{1}{2}\dot{\phi}^2+V(\phi)\right]
\end{equation}
and
\begin{equation}\label{dH}
\dot{H}=-\frac{\kappa^2}{2}(\rho_m+p_m+\dot{\phi}^2),
\end{equation}
where $\kappa^2\equiv 8\pi/M_p^2$, $\rho_m$ and $p_m$ are the
energy density and pressure of the barotropic matter. According to
Eqs.(\ref{ddphi})-(\ref{dH}) and the conservation of energy,
$\rho_m$ satisfies the following continuous equation
\begin{equation}\label{sys1}
\dot{\rho_m}+3H(\rho_m+p_m)=-2\left(\frac{\beta}{M_P}\right)^2\phi\dot{\phi}T^4,
\end{equation}
 and $p_m=(\gamma-1)\rho_m$, where $\gamma$ is a
constant, $0\leq \gamma\leq2$, such as radiation($\gamma=4/3$) or
dust ($\gamma=1$). It is clear that when the thermal coupling
parameter $\beta$ becomes zero, the equations
(\ref{ddphi})-(\ref{sys1}) will return to those of the standard
one scalar field quintessence scenario. In current situation, both
quintessence and barotropic matter are not conserved, but the
overall energy is conserved. The energy density ratio of
quintessence and barotropic matter satisfies the following
conditions
\begin{equation}
\dot{r}\equiv
\dot{\left(\frac{\rho_m}{\rho_{\phi}}\right)}=r\left(\frac{Q}{\rho_m}+\frac{Q}{\rho_{\phi}}
-3H\gamma+3H\gamma_{\phi}\right),
\end{equation}
where the parameter $Q=2(\beta/
M_p)^2\phi\dot{\phi}\rho_{\gamma,0}/ a^4$,
$\gamma_{\phi}=1+w_{\phi}$, in which $w_{\phi}$ is the so-called
parameter of state of quintessence which is defined by
$w_{\phi}=p_{\phi}/\rho_{\phi}$, and
\begin{equation}
p_{\phi}=\frac{1}{2}\dot{\phi}^2-V(\phi),
\end{equation}
\begin{equation}
\rho_{\phi}=\frac{1}{2}\dot{\phi}^2+V(\phi).
\end{equation}

Rewriting the equations (\ref{ddphi}) and (\ref{sys1}), we have
\begin{equation}
\dot{\rho_m}+3H(w_m^{eff}+1)\rho_m=0,
\end{equation}
\begin{equation}
\dot{\rho_{\phi}}+3H(w_{\phi}^{eff}+1)\rho_{\phi}=0,
\end{equation}
where
$w_m^{eff}=\gamma-1+\frac{2\beta^2\phi\dot{\phi}\rho_{\gamma,0}}{3M_p^2Ha^4\rho_m}$
and
$w_{\phi}^{eff}=w_{\phi}-\frac{2\beta^2\phi\dot{\phi}\rho_{\gamma,0}}{3M_p^2Ha^4\rho_m}$,
respectively. Now, introducing the following variables:
\begin{eqnarray}\label{defx}
x&=&\frac{\kappa\dot{\phi}}{\sqrt{6}H},\\
\label{defy}
y&=&\frac{\kappa\sqrt{V(\phi)}}{\sqrt{3}H},\\
\label{defz}
z&=&\frac{\kappa}{M_p}\frac{\phi}{\sqrt{3}H}\frac{\sqrt{\rho_{r0}}}{a^2},\\
\label{defxi}
 \xi &=&\frac{\sqrt{6}}{\kappa \phi},\\
\lambda
&=&-\frac{\sqrt{6}V'(\phi)}{\kappa V(\phi)},\\
\Gamma&=&\frac{V(\phi)V''(\phi)}{V'^2(\phi)},\\
N&=&\ln a,
\end{eqnarray}
the equation system(\ref{ddphi})-(\ref{sys1}) becomes the
following
 system:
\begin{eqnarray}\label{auto1}
\frac{dx}{dN}&=&\frac{3}{2}x[\gamma(1-x^2-y^2)+2x^2]-(3x+\beta^2z^2\xi-\frac{1}{2}\lambda
y^2),\nonumber\\
\frac{dy}{dN}&=&\frac{3}{2}y[\gamma(1-x^2-y^2)+2x^2]-\frac{1}{2}\lambda
xy,\nonumber\\
\frac{dz}{dN}&=&\frac{3}{2}z[\gamma(1-x^2-y^2)+2x^2]-2z+xz\xi,\\
\frac{d\xi}{dN}&=&-x\xi^2,\nonumber\\
\frac{d\lambda}{dN} &=&-x\lambda^2(\Gamma-1). \nonumber
\end{eqnarray}
The energy density parameter  and the equation-of-state parameter
of quintessence, $\Omega_{\phi}$ and $w_{\phi}$, satisfy the
constraint equation
\begin{equation}
\Omega_{\phi}\equiv
\frac{\kappa^2\rho_{\phi}}{3H^2}=1-\frac{\kappa^2\rho_m}{3H^2}=x^2+y^2,
\end{equation}
and $w_{\phi}=(x^2-y^2)/(x^2+y^2)$, respectively. Since the energy
density of the barotropic fluid $\rho_m$ is
semi-positive-definite, any cosmological model can be represented
as a trajectory in the phase space that is bounded within the unit
disc, i.e. $0\le x^2+y^2\le 1$. From the definitions of these new
variables, It is not difficult to reduce the effective parameter
of state of quintessence to
\begin{equation}
w_{\phi}^{eff}=w_{\phi}-\frac{2}{3}\frac{\beta^2xz^2\xi}{1-x^2-y^2}.
\end{equation}

To be concrete, we consider the quintessence with an exponential
potential energy density, i.e.,
\begin{equation} \label{potential1}
 V(\phi)=V_0\exp(-\lambda_0 \kappa \phi)
 \end{equation}
 where the parameter $V_0$ and $\lambda_0$ are two constants.
 Exponential potentials have been studied extensively in various
 situations, and these are of interest for two main reasons.
 Firstly, they can be derived from a good candidate of fundamental
 theory for such being string/M-theory; Secondly, the motion
 equations can be written as an autonomous system in the
 situation.  The author of an earlier work \cite{Halliwell}
 considered a scalar field with a single exponential potential in
 a homogeneous and isotropic universe. Subsequently, it was
 extended to include barotropic matter \cite{Copeland,Hoogen} and
 multiple scalar fields \cite{Li2}, and generalized for
 anisotropic universe \cite{Billyard}. According to the definitions, the parameters
$\lambda$ and $\Gamma$ both become constants and are equal to
$\lambda_0$ and $1$, respectively. Under the circumstance, the
equations (\ref{auto1}) constitute an autonomous system as
follows,
\begin{eqnarray}\label{auto2}
\frac{dx}{dN}&=&\frac{3}{2}x[\gamma(1-x^2-y^2)+2x^2]-(3x+\beta^2z^2\xi-\frac{1}{2}\lambda_0
y^2),\nonumber\\
\frac{dy}{dN}&=&\frac{3}{2}y[\gamma(1-x^2-y^2)+2x^2]-\frac{1}{2}\lambda_0
xy,\nonumber\\
\frac{dz}{dN}&=&\frac{3}{2}z[\gamma(1-x^2-y^2)+2x^2]-2z+xz\xi,\\
\frac{d\xi}{dN}&=&-x\xi^2.\nonumber
\end{eqnarray}

From the equation of state of matter (\ref{sys1}) and Eqs.
(\ref{defx})-(\ref{defxi}), we obtain that
\begin{equation}
\frac{d\rho_m}{dN}+3\gamma \rho_m+2V_0\beta^2\frac{
 xz^2\xi}{y^2}e^{-\sqrt{6}\lambda_0/\xi}=0,
\end{equation}
where we have chosen the exponential potential (\ref{potential1}).
$\rho_m$ can be directly expressed as
\begin{equation}
\rho_m(N)=e^{-3\gamma N}\left(\rho_{m,0}+2V_0\beta^2\int_0^N\frac{
 xz^2\xi}{y^2}e^{3\gamma N-\sqrt{6}\lambda_0/\xi}dN\right),
\end{equation}
where the parameter $\rho_{M,0}$ is the present value of energy
density of matter. It is easy to find that in the case of $\beta=0$,
the energy density of matter will evolve in the manner of $\rho_m
\sim a^{-3}$ for $\gamma =1$ as usual.

When barotropic matter is under consideration  and/or the equations
of motion are too difficult to solve analytically, phase space
methods become particularly useful, because numerical solutions with
random initial conditions usually do not expose all the interesting
properties. In table \ref{table1}, we list the critical points,
conditions for their existence and the cosmological parameters
there.

To investigate the stability of these critical points, we can write
the variables near these points ($x_c$,$y_c$,$z_c$,$\xi_c$) in the
form $x=x_c+\eta_1$,$y=y_c+\eta_2$, $z=z_c+\eta_3$ and
$\xi=\xi_c+\eta_4$ with $\eta_i$,$ i=1,2,3,4$, the perturbations of
the variables about the critical points to the first order. This
leads to the equation of motion,
\begin{equation}
\textbf{U}'=\textbf{A}\cdot\textbf{U}
\end{equation}
where the 4-column vector $\textbf{U}=(\eta_i)^{T}$, $i=1,2,3,4$
represents the perturbations of the variables and $\textbf{A}$ is a
constant $4\times 4$ matrix. For stability we require the all 4
eigenvalues of $\textbf{A}$ to be negative.  For the critical points
listed in table 1, we find the eigenvalues of the linear
perturbation matrix for different stable critical points, see table
\ref{table2}.

Critical points of type A correspond to the scalar field dominated
solutions ($\Omega_{\phi}=1$), which exist for sufficiently flat
potentials, $\lambda^2<24$. Moreover, for the barotropic index
$\gamma> \lambda^2/18$, they are stable to the extent that this kind
of solutions represent late-time attractors in the presence of a
barotropic fluid. Note that type E is a special case of type A,
i.e., $\lambda=0$.

Critical points of type B correspond to another kind of late-time
attractor where neither the quintessence field nor the barotropic
fluid completely dominates the evolution of the universe. They are
known as scaling solutions where the energy density of quintessence
is proportionated to that of barotropic fluid at late time with
$\Omega_{\phi}=18\gamma/\lambda^2$. It is remarkable that the
barotropic index $\gamma$ is required to be no more than $4/3$ for
the sake of stability.

Critical points of type C denote the barotropic fluid dominated
solutions where $\Omega_{\phi}=0$. They are unstable for all
reasonable values of $\gamma$ and $\lambda$ (i.e. however steep the
potential). The ones of type D are the so-called kinetic-dominated
solutions where the late-time evolution of the universe is dominated
by the kinetic energy of quintessence with a stiff equation of
state, i.e. $w_{\phi}=1$. These solutions are unstable as one
expects.

\begin{table*}
\begin{center}
\begin{tabular}{|c| c |c| c| c| c |}
  \hline\hline
  Type & critical points
  ($x_c$,$y_c$,$z_c$,$\xi_c$)  & $\lambda$  & $\Omega_{\phi}$ & $w_{\phi}$ & stability\\
\hline
   A&$\frac{\lambda}{6}$, $\pm\sqrt{1-\frac{\lambda^2}{36}}$, 0, 0  & any & 1 & $-1+\frac{\lambda^2}{18}$&
   stable for $\lambda^2<24$ and $\gamma>\frac{\lambda^2}{18}$ \\
  B&$3\frac{\gamma}{\lambda}$, $\pm3\sqrt{\frac{\gamma(2-\gamma)}{\lambda^2}}$,0,0 & $\lambda\neq 0$& $\frac{18\gamma}{\lambda^2}$ & $\gamma-1$&
   stable for $\gamma<\frac{4}{3}$ and $\lambda^2>18\gamma$\\
  C&0,0,0,any & any & 0 & undefined & unstable \\
 D&$\pm 1$,0,0,0 & any &1 &1 & unstable\\
 E&$0$,$\pm 1$,0,0 & $\lambda=0$ &1 &-1& stable\\
   \hline\hline
\end{tabular}
\caption{The critical points and their physical properties for
 models with exponential potential.}\label{table1}
\end{center}
\end{table*}

\begin{table*}
\begin{center}
\begin{tabular}{|c | c | c |}
  \hline\hline
 Type &Critical points ($x_c$, $y_c$, $z_c$, $\xi_c$) & eigenvalues \\
  \hline
  A&$\frac{\lambda}{6}$, $\pm\sqrt{1-\frac{\lambda^2}{36}}$, 0, 0  &
  $0, -3+\frac{\lambda^2}{12},-3\gamma+\frac{\lambda^2}{6},-2+\frac{\lambda^2}{12}$\\
  B&$3\frac{\gamma}{\lambda}$, $\pm3\sqrt{\frac{\gamma(2-\gamma)}{\lambda^2}}$,0,0&
  $0,
  \frac{3}{4}\left[(\gamma-2)-\sqrt{(2-\gamma)\left(2-9\gamma+\frac{144\gamma^2}{\lambda^2}\right)}\right]$\\
   & &$-2+\frac{3\gamma}{2},\frac{3}{4}\left[(\gamma-2)+\sqrt{(2-\gamma)\left(2-9\gamma+\frac{144\gamma^2}{\lambda^2}\right)}\right]$\\
  C&0,0,0,any & $0, -3+\frac{3\gamma}{2}, \frac{3\gamma}{2}, -2+\frac{3\gamma}{2}$\\
 D&$\pm 1$,0,0,0 &$0, 6-3\gamma, 1, 3\mp \frac{\lambda}{2}$\\
 E&$0$,$\pm 1$,0,0 &$0, -3, -2, -3\gamma$\\
  \hline\hline
\end{tabular}
\caption{The eigenvalues of the critical points for the exponential
potential.}\label{table2}
\end{center}
\end{table*}

\begin{figure}
\begin{center}
\epsfig{file=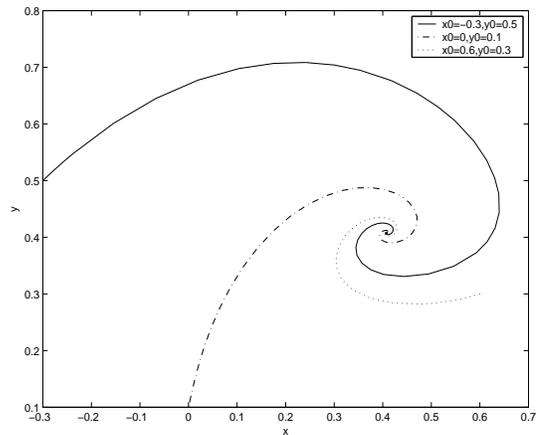,height=2.3in,width=2.8in} \caption{The
portrait of scaling behavior of the quintessence with thermal
interaction for exponential models $V(\phi)=V_0\exp(-\lambda_0
\kappa \phi)$, where $\lambda_0=25$ and the parameter $\gamma
=1$.}\label{fig1}
\end{center}
\end{figure}

As shown in table \ref{table1}, all the critical points exist when
and only when $z=0$, that is to say that $\beta$ does not appear
in the final expressions that determine the critical points. This
means that with the evolution of the universe the thermal coupling
(\ref{eq1}) between the quintessence and matter always goes
asymptotically to vanish away. Fig.\ref{fig1} shows a scaling
solution that the quintessence goes towards an attractor.

\begin{figure}
\begin{center}
\epsfig{file=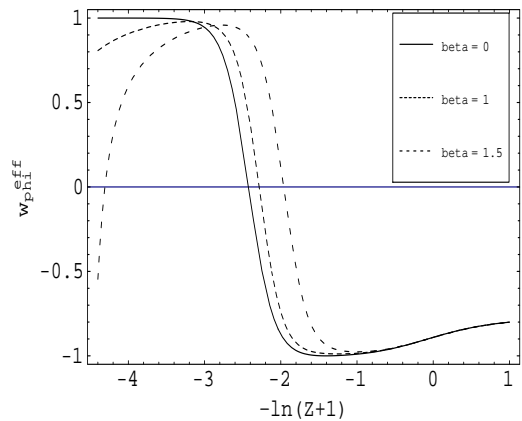,height=2.3in,width=2.8in} \caption{Evolution
of the effective parameter of state $w_{\phi}^{eff}$ for
$\beta=1.5$, $\beta=1$ and $\beta=0$, respectively, where we let
$\gamma =1$, $\lambda_0=2$ and $z$ in the label of horizontal axis
denote red shift.}\label{fig4}
\end{center}
\end{figure}

In the above, we have studied the phase space of scalar field with
thermal interactions to cosmic matter in a flat FRW cosmological
background. The critical points indicate that these stable attractor
phases corresponding to the vanishing of the thermal interaction
between the quintessence and the barotropic matter. However, what
will be the effects of the thermal interactions in the
circumstances? In fact, the introduction of the interacting term
will change the evolutionary tracks of the dynamical systems, which
was not well manifested by the above qualitative analysis.
Therefore, we study their dynamical evolution numerically and
compare the results with the cases of no coupling quintessence
scenario. In Fig.\ref{fig4}, the evolution of equation-of-state
parameter of quintessence is manifested. Obviously, the effective
equation-of-state parameter of quintessence $w_{\phi}^{eff}$ that
incorporate the effect of the thermal coupling between quintessence
and matter is smaller than that of quintessence without coupling to
the background matter.

\begin{figure}
\begin{center}
\epsfig{file=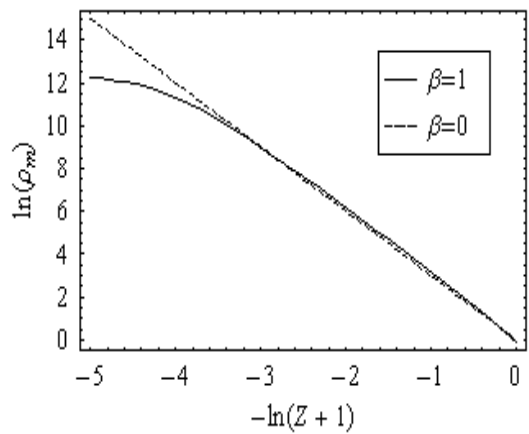,height=2.3in,width=2.8in} \caption{Evolution
of the energy density of barotropic matter $\rho_m$ for $\beta=1$
and $\beta=0$, respectively, where we let $\rho_{m,0}=1$, $\gamma
=1$, $\lambda_0=2$ and $z$ in the label of horizontal axis denote
red shift.}\label{fig3}
\end{center}
\end{figure}

In conclusion, we have investigated the cosmological dynamics of
scalar field, quintessence, that couple to the background matter
with thermal interactions in exponential potential. The conditions
for the existence and stability of various critical points are
obtained by phase space analysis. We find that incorporating the
effect of thermal coupling between the quintessence and background
matter does not qualitatively alter the late-time evolution of the
components in the universe. Just as the ordinary situation that
there is no direct coupling between quintessence and background
barotropic matter \cite{Copeland}, for the parameters
$\lambda^2<18\gamma$, the universe is dominated by quintessence at
late time; and for $\lambda^2>18\gamma$, quintessence does not
entirely dominate the universe but remains a fixed fraction of the
total matter at late time. However, we also find that the effects
of thermal interaction make the equation-of-state parameter of
quintessence become lower than the one in the standard
quintessence scenario of dark energy at the early time and depress
the energy density of matter simultaneously, see Fig.\ref{fig3}.
In other words, large coupling to relic particles such as
neutrinos can be ruled out as they lead to a problematic
equation-of-state parameter, which is consistent with Ref.
\cite{Hsu}. The fact that  $\phi$ is probably close to the Planck
energy suggests that more interactions should figure in to the
behavior of quintessence. However, this will possibly lead to a
relic density problem at nucleosynthesis. Finally, it is worth
emphasizing that this thermal interaction may be helpful to
understand the coincidence of $\Omega_m \sim \Omega_{\phi}$.

\section*{Acknowledgments}
This work is supported by Shanghai Municipal Education Commission
No. 04DC28, Shanghai Municipal Science and Technology Commission No.
04dz05905 and National Natural Science Foundation of China under
Grant No. 10473007.

\end{document}